\documentclass[aps,prb,showkeys,preprintnumbers,amsmath,amssymb]{revtex4}
\usepackage{graphicx}
\usepackage{dcolumn}
\usepackage{bm}
\begin{document}
\def\ibd {\it ibid.}
\def\eps{\epsilon}
\def\vep{\varepsilon}
\def\bfk{{\bf k}}
\def\bfk'{{\bf k'}}
\def\bfq{{\bf q}}
\def\bfr{{\bf r}}
\def\bfp{{\bf p}}
\def\pl{\partial}
\def\tl{\tilde}
\def\zbar{{\bar z}}
\def\al{\alpha}
\def\be{\beta}
\def\gam{\gamma}
\def\lam{\lambda}
\def\bfrho{\bf \rho}
\def\nab{\bf \nabla}
\def\Delmu{\Delta \mu}
\def\beq{\begin{equation}}
\def\eeq{\end{equation}}
\def\bea{\begin{eqnarray}}
\def\eea{\end{eqnarray}}
\title{Ground-state energy density of a dilute Bose
gas in the canonical transformation}
\author{Sang-Hoon \surname{Kim},$^1$\footnote{shkim@mmu.ac.kr}
 Chul Koo \surname{Kim},$^{2}$ and Mukunda P. \surname{Das}}
\affiliation{$^1$Division of Liberal Arts, Mokpo National Maritime
University,  Mokpo 530-729, Korea,} \affiliation{$^2$Institute of
Physics and Applied Physics,
 Yonsei University, Seoul 120-749, Korea,}
 \affiliation{$^3$Department of Theoretical Physics, RSPhysSE,
The Australian National University, Canberra, ACT 0200, Australia}
\begin{abstract}
A ground state energy density of an interacting dilute Bose gas
 system is studied in the canonical transformation scheme.
It is shown that the transformation scheme enables us to
 calculate a higher order correction of order $n a^3$
in the particle depletion and ground state energy density
 of a dilute Bose gas system,
which corresponds to the density fluctuation contribution
 from the excited states.
Considering two-body interaction only, the coefficient of $n a^3$
term is shown to be $2(\pi - 8/3)$
 for the particle depletion,
and $16(\pi - 8/3)$ for the ground state energy density.
\end{abstract}
\maketitle

\section{Introduction}
Investigation of an interacting dilute Bose gas has a long history
beginning with the seminal works of Bogoliubov, Beliaev, Lee, Huang
and Yang, Hugenholtz and Pines. This topic has been amply discussed
in standard text books. \cite{huan,abri,fett}
 Despite the
successful demonstration of Bose-Einstein condensation (BEC) in
inhomogeneous system of dilute alkali gases during the last decade,
the dilute interacting Bose fluids in homogeneous system is one of
the fundamental interests still in statistical physics.

 The ground state energy density of the dilute system is expressed
in the following general form: \cite{huge}
\begin{equation}
\frac{E_g}{V} =\frac{2\pi\hbar^2a n^2}{m}\sum_{i=0}^\infty
\sum_{j=0}^{[i/2]} C_{ij} (na^3)^{i/2} \{\ln(n a^3)\}^j, \label{1}
\end{equation}
where $n=N/V$ is the particle density. Since unconfined BEC occurs
for a two-body repulsive interaction, the $s$-wave scattering length
$a$ is positive.
 $[i/2]$ is the Gauss number which takes the integer part of $i/2$,
and $C_{ij}$ are the coefficients in the expansion. Through
elaborate calculations, it is known that \cite{braa}
\begin{equation}
\frac{E_g}{V} = \frac{2\pi\hbar^2a n^2}{m} \left[ 1 +
\frac{128}{15\sqrt{\pi}} (na^3)^{1/2} + \left\{ \frac{8(4 \pi - 3
\sqrt{3})}{3} \ln (n a^3) + \kappa \right\} n a^3 + ...  \right].
\label{2}
\end{equation}
Comparing Eq. (\ref{1}) with Eq. (\ref{2}), we find the coefficients
as $C_{00}=1$, $C_{10}= 128/15\sqrt{\pi}$, and
  $C_{21}= 8(4\pi - 3\sqrt{3})/3$.
Nevertheless, the coefficient of $C_{21}$ was calculated early
 in 1950's,\cite{huge,wu,sawa} the constant $\kappa = C_{20}$
 and other higher order terms did not receive much attention.

Hugenholtz and Pines mentioned that the expansion is not a simple
 power series, rather it involves the logarithm of
expansion parameter $n a^3$.\cite{huge} Although  Fetter and Walecka
gave a rough estimate on the fluctuation contribution from the
condensate in the order of $n a^3$, they stated that the coefficient
$C_{20}$ {\it has never been determined}.\cite{fett}
       In order to get a deeper understanding of the ground state
over the mean field results, it is necessary to calculate
 corrections from the quantum fluctuations.
Earlier derivation of rigorous upper and lower bounds had been
 given by Dyson.\cite{dyso}
Also an improved calculation on the lower bounds has been carried by
Lieb and Yngvason.\cite{lieb}

Some practical calculations of the coefficient $C_{20}$ were carried
out very recently by two groups. Our group obtained  $C_{20}$ from
the fluctuation contribution by using the functional Schr\"{o}dinger
picture approach of two-body interaction.\cite{kim} The
Schr\"{o}dinger picture for many particle systems or relativistic
 quantum field theory is an infinite dimensional extension of quantum
 mechanics and leads into a functional formulation.\cite{hatf,ckkim1,ckkim2}
 However, the result has not been confirmed using a separate
 independent method.
Another contribution by three-body interaction has been calculated
 by Braaten {\it et al.}\cite{braa,braa1}
In their calculation the core potential is necessary to find
$C_{20}$, and they calculated $C_{20}$ as a function of an effective
potential parameter $r_s$.\cite{braa} Therefore, the two results for
$C_{20}$ are independent.

In this paper, we consider the contribution from the condensate
fluctuation, which is independent on the details of the potential
and two-body interaction only. We show that the
 result obtained by the Schr\"{o}dinger  picture
can be obtained by a more standard canonical transformation method.
In the process we obtain the fluctuation correction to the particle
depletion, and  $C_{10}$ in Eq. (\ref{1}) from
  second order ground state expansion.
Also, we will consider the fluctuation contribution from the
condensate, and then calculate the coefficients $C_{20}$ and
$C_{30}$. On the other hand $C_{21}$ which come from three-body
interaction will not be touched.

 This paper is organized as follows:
 In Section II, the interacting Hamiltonian of the dilute Bose
system is expanded up to the terms of our interest.
  In Section III, the canonical transformations are applied twice
to diagonalize the Hamiltonian.
  In Section IV, the particle depletion appearing in the diagonalized
Hamiltonian is obtained.
  In Section V, the fluctuation correction to the ground state energy
 density is obtained as a series of  $n a^3$.
  In Section VI, we summarise and discuss our results.

\section{Many-body dilute Bose gas system}
We start with the standard model Hamiltonian for an interacting Bose
gas given by \cite{fett}
\begin{equation}
\hat{H} = \sum_{ k }  \frac{\hbar^2 k^{2}}{2m} a_{k}^{\dag} a_{k} +
\frac{g}{2V} \sum_{ {{ k_{1} k_{2}}  \atop {k_{3} k_{4} }}  }
  a_{k_{1}}^{\dag}  a_{k_{2}}^{\dag} a_{k_{3}} a_{k_{4}}
  \delta_{k_{1}+k_{2},k_{3}+k_{4}}.
\label{3}
\end{equation}
Here, the constant  matrix element  $g$ should  be determined
  by requiring that $\hat{H}$ produces the two-body scattering
 properties in vacuum.
Since the system is  dilute enough, we assume that the $s$-wave
scattering length $a$ is much less than the
 inter-particle spacing $ n^{1/3}$.
As a result  of the $n a^3 \ll 1$, this satisfies to be a small
expansion parameter. Under this  assumption the  zero momentum
operators $a_{0},  a_{0}^{\dag}$  become nearly classical due to
occurrence
 of a condensate.

 In two-body interaction the interacting part of the Hamiltonian in Eq. (\ref{3}) is
expanded to keep  the terms up to
 order of $N_{0}^{2}, N_{0}$, and $\sqrt{N_0}$
as \cite{kim,maha}
\begin{eqnarray}
{\hat{H}}_{int} &=& \frac{g}{2V} a_{0}^{\dag} a_{0}^{\dag} a_{0}
a_{0}
\nonumber \\
& & + \frac{g}{2V} \sum_{ k \neq 0 }
 \left[ 2( a_{k}^{\dag} a_{0}^{\dag} a_{k} a_{0}
+ a_{-k}^{\dag} a_{0}^{\dag} a_{-k} a_{0} ) + a_{k}^{\dag}
a_{-k}^{\dag} a_{0} a_{0} + a_{0}^{\dag} a_{0}^{\dag} a_{k} a_{-k}
\right]
\nonumber \\
& & + \frac{g}{V} \sum_{ k , q \neq  0 }
 \left[   a_{k+q}^{\dag} a_{0}^{\dag}  a_{k} a_{q}
+ a_{k+q}^{\dag} a_{-q}^{\dag} a_{k} a_{0}  \right]. \label{4}
\end{eqnarray}
The last term which is generally neglected in textbook calculations
originates  from the interactions of particles out of and into the
condensate,
 thus representing the condensation fluctuation.
In Ref. [5] the three-body interaction term $(\psi^* \psi)^3$
contributes to the $C_{20}$. However, in our calculation one
momentum among the ${\bf k}_1,{\bf k}_2,{\bf k}_3,{\bf k}_4$ of
two-body interaction is zero and will contribute to the $C_{20}$.
Then, it makes our contribution different from that of Ref. [5].

In order to handle the contribution from the last term, we
introduce a new  variable $\gamma_{k}$
\begin{equation}
\gamma_{k} = \sum_{q \neq 0}  a_{k+q}^{\dag} a_{q}. \label{5}
\end{equation}
Replacing the operators $a_{0} , a_{0}^{\dag}$  by  $\sqrt{N_0}$,
and applying  the number operator relation $ N_0  = \hat{N} -
\sum_{k \neq   0} ( a_{k}^{\dag} a_{k} +  a_{-k}^{\dag} a_{-k})/2$
to Eq. (\ref{4}), we rewrite the model Hamiltonian  as
\begin{eqnarray}
\hat{H} &=&  \frac{1}{2} n^2 V g
 \nonumber \\
  & & + \frac{1}{2} \sum_{k \neq  0}
\left[ (\epsilon^{0}_k + n g) (a_{k}^{\dag} a_{k} + a_{-k}^{\dag}
a_{-k})
 + n g (a_{k}^{\dag} a_{-k}^{\dag} + a_{k} a_{-k})\right]
 \nonumber \\
 & & + \frac{ng}{\sqrt{N}}
 \sum_{k \neq  0} \gamma_k ( a_k  + a_{-k}^{\dag}  ),
\label{7}
\end{eqnarray}
where  $\epsilon^{0}_k = \hbar^2 k^{2}/2m$. The ground state
properties of the above Hamiltonian is to be calculated using  a
canonical transformation method.

\section{Canonical transformation method}

To diagonalize the above Hamiltonian, we apply the Bogoliubov
transformation of the new  the operators $a_k$ and $a_k^\dagger$ as
follows,\cite{huan}
\begin{eqnarray}
a_k &=& \frac{1}{\sqrt{1 - A_k^2}} (b_k + A_k b_{-k}^\dagger),
\nonumber \\
a_k^\dagger &=& \frac{1}{\sqrt{1 - A_k^2}} (b_k^\dagger + A_k
b_{-k}). \label{21}
\end{eqnarray}
Then, the Hamiltonian in Eq. (\ref{7}) becomes
\begin{eqnarray}
\hat{H} &=&  \frac{1}{2} n^2 V g + \sum_{k\neq
0}\frac{1}{1-A_k^2}\left[(\epsilon_k^0 + n g)A_k^2
 + n g A_k \right]
 \nonumber \\
& & + \frac{1}{2}\sum_{k\neq 0}\frac{1}{1-A_k^2} \left[(\epsilon_k^0
+ n g)(1+ A_k^2) + 2 n g A_k  \right] (b_k^\dagger b_k +
b_{-k}^\dagger b_{-k})
 \nonumber \\
& & + \frac{1}{2}\sum_{k\neq 0}\frac{1}{1-A_k^2}
\left[2(\epsilon_k^0 + n g)A_k +  n g(1+ A_k^2)  \right]
(b_k^\dagger b_{-k}^\dagger + b_{k} b_{-k})
 \nonumber \\
& & + \frac{n g}{\sqrt{N}}\sum_{k \neq 0}
\frac{(1+A_k)}{\sqrt{1-A_k^2}}
 (b_k + b_{-k}^\dagger) \gamma_k.
\label{23}
\end{eqnarray}

$A_k$ is chosen to make the off-diagonal term vanish. That is,
\begin{equation}
2(\epsilon_k^0 + n g)A_k +  n g(1+ A_k^2) = 0, \label{25}
\end{equation}
and  we obtain $A_k(=A_{-k})$ as
\begin{equation}
A_k = \frac{ E_k - (\epsilon_k^0 + ng)}{n g}, \label{27}
\end{equation}
where
\begin{equation}
E_k = \sqrt{(\epsilon_k^0 + n g)^2 - (n g)^2}. \label{29}
\end{equation}
The positive sign was selected in Eq. (\ref{27}) to give positive
excited energy values. Note that $(1+A_k)/(1-A_k)=\epsilon_k^0
/E_k$.

Substituting  $A_k$ into the Hamiltonian in Eq. (\ref{23}), we
obtain the following compact form of the Hamiltonian
\begin{eqnarray}
\hat{H} &=&  \frac{1}{2} n^2 V g + \sum_{k\neq
0}\frac{1}{1-A_k^2}\left[(\epsilon_k^0 + n g)A_k^2
 + n g A_k \right]
 \nonumber \\
& & + \sum_{k\neq 0} E_k b_k^\dagger b_k + \sum_{k\neq 0} G_k (b_k +
b_{-k}^\dagger), \label{31}
\end{eqnarray}
where
\begin{equation}
G_k = \frac{ng}{\sqrt{N}}\frac{(1+A_k)\gamma_k}{\sqrt{1-A_k^2}}.
\label{33}
\end{equation}

We can regard the system as a collection of quantum mechanical
 harmonic oscillators exposed to additional forces given
by linear terms. The linear terms $b_k$ and $b_k^\dagger$ can be
made to eliminate by a linear transformation of the form
\begin{eqnarray}
b_k &=& c_k + \alpha_k,
\nonumber \\
b_k^\dagger &=& c_k^\dagger + \alpha_{-k}, \label{35}
\end{eqnarray}
where the new operators $c_k$ and $c_k^\dagger$ satisfy the bosonic
commutator relations. Note that $\alpha_k^\dagger=\alpha_{-k}$ and
commutes with $c_k$.

Substituting the new operators in Eq. (\ref{35}) into Eq.
(\ref{31}), we can rewrite the Hamiltonian in Eq. (\ref{31}) as
\begin{eqnarray}
\hat{H} &=&  \frac{1}{2} n^2 V g + \sum_{k\neq
0}\frac{1}{1-A_k^2}\left[(\epsilon_k^0 + n g)A_k^2
 + n g A_k \right]
\nonumber \\
& & + \sum_{k\neq 0} \left( E_k c_k^\dagger c_k + E_k
\alpha_k^\dagger \alpha_k + 2 G_k \alpha_k \right)
 \nonumber \\
& & + \sum_{k\neq 0} (E_k \alpha_{-k} + G_k) c_k +  \sum_{k\neq 0}
(E_k \alpha_{k} + G_{-k}) c_k^\dagger. \label{37}
\end{eqnarray}
Here, we choose $\alpha_k = -G_{-k}/E_k$
 to make the linear terms vanish.
Then, the Hamiltonian in Eq. (\ref{37}) is now
 diagonalized in terms of new operators as follows
\begin{eqnarray}
\hat{H} &=&  \frac{1}{2} n^2 V g + \sum_{k\neq
0}\frac{1}{1-A_k^2}\left[(\epsilon_k^0 + n g)A_k^2
 + n g A_k \right]
 \nonumber \\
& & - \sum_{k \neq 0} \frac{G_k G_{-k}}{E_k} + \sum_{k\neq 0} E_k
c_k^\dagger c_k
\nonumber \\
&=& \hat{H}_g + \sum_{k\neq 0} E_k n_k. \label{39}
\end{eqnarray}

Using Eqs. (\ref{27}), (\ref{29}), (\ref{33}), and the chosen
 $\alpha_k$,
 the above Hamiltonian $\hat{H}_g$ produces the general form of
the ground state energy  as
\begin{eqnarray}
E_g &=&  \frac{1}{2} n^2 V g + \frac{1}{2}\sum_{k\neq 0}\left[ E_k -
(\epsilon_k^0 + n g) \right] + \frac{n^2 g^2}{N}\sum_{k\neq 0}
\frac{\epsilon_k^0}{E_k^2}\langle 0|\gamma_k \gamma_{-k}|0 \rangle
\nonumber \\
&\equiv& E_0 + E_1 + E_2. \label{40}
\end{eqnarray}
From the relation of the interaction strength, $g$, \cite{abri}
 \begin{equation}
 \frac{4\pi \hbar^2 a}{m}
 = g - \frac{g^2}{2V}\sum_{k \neq 0}\frac{1}{\epsilon^{0}_k} + ...\, ,
 \label{41}
 \end{equation}
we obtain  the ground state energy density of the first two terms as
\begin{equation}
\frac{1}{V}(E_0 + E_1)
 =  \frac{2\pi\hbar^2 a n^2 }{m}
 \left[ 1+ \frac{128}{15\sqrt{\pi}} (na^3)^{1/2} \right].
\label{42}
\end{equation}
  Note that $\sum_k$ is
replaced with $V\int d^3k / (2\pi)^3$ during the calculation. The
first two terms, $E_0$ and $E_1$ in the ground state
 energy are exactly  same as those obtained through conventional
methods in  the  literature.\cite{huan,abri,fett}

\section{The particle depletion}

Accurate determination of the fractional depletion of the zero
momentum state for an interacting gas is increasingly needed with
the advance of the experimental technique on the BEC states.
 The particle distribution of the dilute system is obtained
from Eq. (\ref{21}) and (\ref{35}) as
\begin{eqnarray}
n_k &=& \langle 0| a_k^\dagger a_k |0 \rangle
 \nonumber \\
&=& \frac{A_k^2}{1-A_k^2} + \frac{(1+A_k)^2}{1-A_k^2} \langle 0|
\alpha_{-k} \alpha_k |0 \rangle
\nonumber \\
&=& \frac{1}{2} \left( \frac{\eps_k^0 + n g}{E_k} - 1 \right) +
\frac{n^2 g^2}{N}\frac{{\eps_k^0}^2}{E_k^4} \langle 0| \gamma_k
\gamma_{-k} |0 \rangle. \label{43}
\end{eqnarray}

The dominant part of the unknown quantity
 $\langle  \gamma_k \gamma_{-k}  \rangle$
is obtained by iteration method. Applying the definition of $
\gamma_k$ in Eq. (\ref{5}), it is written as
\begin{equation}
\langle 0| \gamma_k \gamma_{-k} |0 \rangle = \langle 0|
\sum_{q,q'\neq 0} a_{k+q}^\dagger a_{q} a_{-k+q'}^\dagger a_{q'} |0
\rangle. \label{44}
\end{equation}
     It is known that the dominant contribution arises
when a particle interacts with itself and it belongs to the term
where ${\bf q'} = {\bf k} + {\bf q}$. Then, we separate the dominant
contribution into two parts ${\bf q}={\bf q'}$ and ${\bf q} \neq
{\bf q'}$ again.
\begin{eqnarray}
\langle 0| \gamma_k \gamma_{-k} |0 \rangle &=&\langle 0|
\sum_{q'=k+q} a_{k+q}^\dagger a_{q} a_{-k+q'}^\dagger a_{q'} |0
\rangle + \langle 0| \sum_{q'\neq k+q} a_{k+q}^\dagger a_{q}
a_{-k+q'}^\dagger a_{q'} |0 \rangle
\nonumber \\
&=& \langle 0| \sum_{q=q'\neq 0} a_{q'}^\dagger a_{q} a_{q}^\dagger
a_{q'} |0 \rangle + \langle 0| \sum_{q\neq q' \neq 0} a_{q'}^\dagger
a_{q} a_{q}^\dagger a_{q'} |0 \rangle
 \nonumber \\
 & & + ({\rm k~dependent~minor~terms}).
\label{45}
\end{eqnarray}
The term for ${\bf q} \neq {\bf q'}$ vanishes using the argument of
random phase approximation.

  Within this approximation we obtain the dominant part of
$\langle \gamma_k \gamma_{-k} \rangle $ as
\begin{eqnarray}
\langle 0| \gamma_k \gamma_{-k} |0 \rangle &=& \sum_{q\neq 0} n_q^2
+ \dots
\nonumber \\
&=& \sum_{q\neq 0} \left\{ \frac{1}{2} \left( \frac{\eps_q^0 + n
g}{E_q} - 1 \right)\right\}^2 + \dots
 \nonumber \\
&=& N \left( \pi - \frac{8}{3} \right) p + \dots. \label{46}
\end{eqnarray}
The new variable $p=\sqrt{n a^3/\pi}$ ($0 < p \ll 1$) accounts for
the diluteness of the system and will be used for the expansion.

     When we back $\langle \gamma_k \gamma_{-k} \rangle $
in Eq. (\ref{46}) into Eq. (\ref{43}) by iteration method, we obtain
a new $n_k$ as
\begin{equation}
n_k = \frac{1}{2} \left( \frac{\eps_k^0 + n g}{E_k} - 1 \right) +
\frac{n^2 g^2 {\eps_k^0}^2 }{E_k^4} \frac{ {\eps_k^0}^2 }{E_k^4}
\left( \pi - \frac{8}{3} \right) p + \dots. \label{47}
\end{equation}
Putting the new $n_k$ into Eq. (\ref{46}), we obtain
 $ \langle \gamma_k \gamma_{-k} \rangle$ as
\begin{eqnarray}
 \langle 0| \gamma_k \gamma_{-k} |0 \rangle
&\simeq& \sum_{q \neq 0} \left\{ \frac{1}{2}\left( \frac{\eps_q^0 +
n g}{E_q} - 1 \right) + \frac{n^2 g^2{\eps_q^0}^2}{E_q^4} \left( \pi
- \frac{8}{3} \right) p \right\}^2
\nonumber \\
&=&N \left[\left( \pi - \frac{8}{3}\right)p + 2\left(\pi -
\frac{8}{3}\right)\left( \frac{10}{3}-\pi\right)p^2 + 2\pi \left(\pi
- \frac{8}{3}\right)^2 p^3 \right]. \label{48}
\end{eqnarray}
We see the dominant part of $ \langle \gamma_k \gamma_{-k} \rangle$
is $k$-independent. Therefore, $E_2$ in Eq. (\ref{40}) can be
written as
\begin{equation}
E_2  =  \frac{n^2 g^2  }{ N}  \langle 0| \gamma_k \gamma_{-k} |0
\rangle \sum_{q\neq 0} \frac{1}{\epsilon_q^0 +2 n g}.
 \label{50}
\end{equation}

Finally substituting Eq. (\ref{48}) into Eq. (\ref{43}), we obtain
 the particle depletion from the zero momentum condensate  as
\begin{eqnarray}
\frac{N-N_0}{N} &=& \frac{1}{N}\sum_{q \neq 0} n_q
 \nonumber \\
 &=&\frac{1}{2N}\sum_{q \neq 0}\left( \frac{\eps_q^0 + n g}{E_q}
- 1\right)
\nonumber \\
& & +  \frac{n^2 g^2}{N}\sum_{q \neq 0}\frac{{\eps_q^0}^2}{E_q^4}
\left[\left( \pi - \frac{8}{3}\right) p + 2\left(\pi -
\frac{8}{3}\right) \left( \frac{10}{3}-\pi \right)p^2 + \dots
\right]
\nonumber\\
&=& \frac{8}{3}\sqrt{\frac{n a^3}{\pi}} + 2\left( \pi -
\frac{8}{3}\right) n a^3 + \dots. \label{61}
\end{eqnarray}

We obtained the next order correction term of the particle depletion
 beyond the textbook result as $2(\pi - 8/3)n a^3$.\cite{huan,abri,fett}
  If we repeat the above procedure, we can obtain
the higher order terms too within the given approximation scheme.
 In  the following section, we discuss the next order correction to the
ground state  energy density which originates from the fluctuation
contribution of the condensate.

\section{The higher order correction to the ground state energy density}

This approximation of $ \langle \gamma_k \gamma_{-k} \rangle$
 also gives us a simple
 result for the condensation fluctuation correction $E_2$ in  Eq. (\ref{50})
to the ground state energy density, but the integral carries an
ultraviolet divergence.
 The same divergence in $E_1$ was handled through the expression
of the  scattering length $a$ to order of $g^2$ as given in  Eq.
(\ref{41}). However, for the calculation of $E_2$ such a simple
cancellation scheme
 is not possible, because it is now necessary to expand $a$ up to $g^3$.
Therefore, we resort to an alternative cutoff procedure
 called {\it minimal subtraction renormalization scheme}.
The justification for this scheme to the current problem is given in
detail in Ref. [5].

In this scheme the ultraviolet divergence in Eq. (\ref{50}) can be
regularized by imposing a cutoff procedure in the $k$ space. With
the cutoff $|{\bf k}| < \Lambda$ where $\Lambda$ is an arbitrary
separation between short distance effect, the divergence can be
handled by introducing a counter term in the Hamiltonian.
 That is, linear, quadratic, and other power ultraviolet
 divergences are removed as part of the regularization scheme by subtracting
the appropriate power of $k$ from the momentum space integrated. In
this case $1/k^2$ is subtracted from the integral. The justification
for this procedure is that the terms that are subtracted are
dominated by short distances and can be canceled by counter terms in
the Hamiltonian. Also, we note that the result is independent of the
potential used, since the fluctuation contribution from the
condensate considered here is a generic feature for any
potential.\cite{fett}

Following this prescription, we obtain $E_2$ in a dimensionless form
\begin{equation}
\frac{E_2}{E_c} = - \frac{2 g }{N V} \langle 0| \gamma_k \gamma_{-k}
|0 \rangle \sum_{q \neq 0} \left( \frac{1}{\epsilon_q^0 + 2 n g} -
\frac{1}{\epsilon_q^0} \right), \label{81}
\end{equation}
where $E_c = n^2 g V/2 = 2\pi\hbar^2 a n^2 V/m$. The summation part
in Eq. (\ref{81}) is obtained by integration independently as
\begin{eqnarray}
 \sum_{q \neq 0}
\left( \frac{1}{\epsilon_q^0 + 2 n g} - \frac{1}{\epsilon_q^0}
\right) &=& -\frac{4 m^2 n g}{\pi^2 \hbar^4}\int_0^\infty
\frac{dq}{q^2 + \frac{4 m n g}{\hbar^2}}
\nonumber \\
&=& -\frac{2 m V}{\hbar^2 a}p \label{82}
\end{eqnarray}

Substituting Eqs. (\ref{41}), (\ref{47}) and (\ref{82}) into Eq.
(\ref{81}), this approximation gives the result for the condensation
fluctuation correction, $E_2$,
\begin{eqnarray}
\frac{E_2}{E_c} =  16 \pi \left(\pi-\frac{8}{3}\right) p^2 + 32 \pi
\left(\pi-\frac{8}{3}\right) \left( \frac{10}{3}-\pi \right) p^3 +
{\cal O}(p^4). \label{84}
\end{eqnarray}
Thus, the present calculation gives  $C_{20} = 16 (\pi-8/3)$, and
$C_{30} = (32/\sqrt{\pi}) (\pi-8/3) (10/3-\pi)$. The $C_{20}$ is
exactly the same as one from our previous functional Schr\"{o}dinger
picture method.\cite{kim} The higher order contribution to the
ground state energy in Eq. (\ref{84}) comes from the interactions of
particles out of and into the condensate. This term represents the
fluctuation
 contribution from the excited states and disappears if the number of
 condensate particles is conserved strictly.

\section{Summary and discussions}

In this paper we revived the ground state energy density calculating
of an interacting dilute Bose gas. Hugenholtz and Pines have given a
general form of the series expansion in terms of $\sqrt{n a^3}$.
Till now only the first few coefficients are known. In view of the
fundamental interest of
 homogeneous dilute interacting Bose gas,
it is desirable to know about the coefficients of higher order
expansion.

There are at least two possible sources to give contributions for
the unknown coefficients of $n a^3$ terms. One is the fluctuation
from condensation and produces a universal result. The other is from
the three-body interaction and not universal because it depends on
the radial potential. Here we used the standard canonical
transformation method of two-body interaction. We obtained the next
order terms in the particle depletion and the ground state energy
density coming from the condensate fluctuation. They are $2(\pi -
8/3)$ and $16(\pi - 8/3)$ respectively. Furthermore, the coefficient
of $(n a^3)^{3/2}$ term is $(32/\sqrt{\pi}) (\pi-8/3) (10/3-\pi)$ in
this approximation.

\acknowledgments
Professor H.S. Noh participated actively in this project.
Unfortunately he passed away last spring. With heavy heart the
authors dedicate this paper to the departed soul for his excellent
academic accomplishments.
S.-H. Kim  acknowledges that this work was
supported by the Korean Research Foundation
Grants(KRF-2005-013-C0021).
C.K. Kim acknowledges support from Korea
Science and Engineering Foundation through Center for Strongly
Correlated Materials Research, SNU(2005).

{}
\end{document}